\documentclass[prb,aps,showpacs,twocolumn,unsortedaddress,superscriptaddress]{revtex4-1}
\usepackage{graphics, bm, psfrag, amsmath, amssymb, epsfig, grffile, float}
\usepackage{subfigure}
\usepackage{color}

\begin{document}

\title{Odd-frequency Superconductivity in Driven Systems}

\author{Christopher Triola}
\affiliation{Nordita, Roslagstullsbacken 23, SE-106 91 Stockholm, Sweden}
\affiliation{Center for Quantum Materials (CQM), KTH and Nordita, Stockholm, Sweden}
\author{Alexander V. Balatsky}
\affiliation{Institute for Materials Science, Los Alamos National Laboratory, Los Alamos New Mexico 87545, USA}
\affiliation{Nordita, Roslagstullsbacken 23, SE-106 91 Stockholm, Sweden}
\affiliation{Center for Quantum Materials (CQM), KTH and Nordita, Stockholm, Sweden}


\begin{abstract}
We show that Berezinskii's classification of the symmetries of Cooper pair amplitudes holds for driven systems even in the absence of translation invariance. We then consider a model Hamiltonian for a superconductor coupled to an external driving potential and, treating the drive as a perturbation, we investigate the corrections to the anomalous Green's function, density of states, and spectral function. We find that in the presence of an external drive the anomalous Green's function develops terms that are odd in frequency and that the same mechanism responsible for these odd-frequency terms generates additional features in the density of states and spectral function. 
\end{abstract}


\maketitle

%

\section{Introduction}

Originally posited in the context of $^3$He by Berezinskii\cite{Berezinskii1974} and then later for superconductivity\cite{BalatskyPRB1992} odd-frequency (odd-$\omega$) pairing offers a theoretical means for interacting Fermions to avoid strong on-site repulsion by pairing Fermions at different times. An intriguing consequence of odd-$\omega$ pairing is that it introduces alternatives to the conventional classification of Fermionic pairs as either spin triplet odd-parity or spin singlet even-parity. Given the inherently dynamical nature of odd-$\omega$ pairing it seems natural to discuss the possibility of its emergence in systems that possess explicit time-dependence, however we are unaware of any previous study with this focus. The purpose of this paper is to elucidate the fact that odd-$\omega$ superconductivity is a concept generalizable to driven systems, even in the absence of translation invariance and to discuss the conditions under which coupling a conventional superconductor to a time-dependent potential should give rise to odd-$\omega$ pair amplitudes. Additionally, we demonstrate explicitly that in the presence of a time-dependent and spatially nonuniform electric field a conventional superconductor will develop odd-$\omega$ pair amplitudes. The emergence of these odd-$\omega$ amplitudes is then shown to be related to phenomena measurable using scanning tunneling microscopy (STM) and photoemission spectroscopy. 

Previous research on odd-$\omega$ pairing has been devoted to the study of the thermodynamic stability of odd-$\omega$ phases\cite{heid1995thermodynamic,solenov2009thermodynamical,kusunose2011puzzle} as well as the identification of systems in which odd-$\omega$ pairing could be induced including: ferromagnetic - superconductor heterostructures \cite{BergeretPRL2001,bergeret2005odd,yokoyama2007manifestation,houzet2008ferromagnetic,EschrigNat2008,LinderPRB2008,crepin2015odd} with recent experimental indications of its realization at the interface of Nb thin film and epitaxial Ho\cite{di2015signature}, certain kinds of topological insulator - superconductor systems \cite{YokoyamaPRB2012,Black-SchafferPRB2012,Black-SchafferPRB2013,TriolaPRB2014}, certain normal metal - superconductor junctions due to broken symmetries\cite{tanaka2007theory,TanakaPRB2007,LinderPRL2009,LinderPRB2010_2,TanakaJPSJ2012}, and multiband superconductors due to finite interband hybridization\cite{black2013odd,komendova2015experimentally}. Additional work has investigated the emergence of proximity-induced odd-$\omega$ pairing in generic two-dimensional bilayers coupled to conventional $s$-wave superconductors\cite{parhizgar_2014_prb} and in two-dimensional electron gases coupled to superconductor thin films\cite{triola2016}. Furthermore, the concept of odd-$\omega$ order parameters can be generalized to charge and spin density waves\cite{pivovarov2001odd,kedem2015odd} and Majorana fermions\cite{huang2015odd}, demonstrating the ubiquity of the odd-$\omega$ class of ordered states. 

\section{Odd-frequency Pairing in Driven Systems}

To establish the relevance of odd-$\omega$ pairing in driven systems we start by noting that the anomalous Green's function of a superconductor can be defined as
$F_{\sigma_1,\alpha_1;\sigma_2,\alpha_2}(\textbf{r}_1,t_1;\textbf{r}_2,t_2)=-i\langle T \psi_{\sigma_1,\alpha_1,\textbf{r}_1,t_1}\psi_{\sigma_2,\alpha_2,\textbf{r}_2,t_2}\rangle$
where $T$ is the time-ordering operator, and $\psi_{\sigma,\alpha,\textbf{r},t}$ annihilates an electron at time $t$, position $\textbf{r}$, with spin $\sigma$, and orbital degrees of freedom described by $\alpha$. Using the definition of time-ordering for Fermions it is straightforward to show that: $F_{\sigma_1,\alpha_1;\sigma_2,\alpha_2}(\textbf{r}_1,t_1;\textbf{r}_2,t_2)=-F_{\sigma_2,\alpha_2;\sigma_1,\alpha_1}(\textbf{r}_2,t_2;\textbf{r}_1,t_1)$. This relation tells us that the wavefunction describing the Cooper pairs, $\Psi$, must obey $\mathcal{S}\mathcal{T}\mathcal{O}\mathcal{P}\Psi=-\Psi$ where: $\mathcal{S}$ acts on spin ($\sigma_1\leftrightarrow\sigma_2$); $\mathcal{T}$ reverses time coordinates ($t_1\leftrightarrow t_2$); $\mathcal{O}$ interchanges orbital degrees of freedom ($\alpha_1\leftrightarrow\alpha_2$); and $\mathcal{P}$ is the spatial parity operator ($\textbf{r}_1\leftrightarrow\textbf{r}_2$). Using this property of $\Psi$ together with the fact that all four transformations are involutory, the possible symmetries of the Cooper pair wavefunction may be divided into 8 different classes based on how they transform under $\mathcal{S}$, $\mathcal{T}$, $\mathcal{O}$, and $\mathcal{P}$, see Table~\ref{table:classification}. 
\begin{center}
\begin{table}[htb]
\begin{tabular}{c || c | c | c | c | c | c | c | c}
Spin & -1 & 1  & 1  & 1  & 1  & -1 & -1 & -1 \\ 
\hline
Time & 1  & 1  & 1  & -1 & -1 & -1 & -1 & 1 \\
\hline
Orbital & 1  & 1  & -1 & 1  & -1 & -1 & 1  & -1 \\
\hline
Parity & 1  & -1 & 1  & 1  & -1 & 1  & -1 & -1 
\end{tabular}
\caption{Classification of the 8 allowed symmetry classes for the superconducting order parameter (Cooper pair wavefunction).}
\label{table:classification}
\end{table}
\end{center}

Fourier transforming the anomalous Green's function with respect to the average and relative position and time it can be shown that:
\begin{equation}
F_{\sigma_1,\alpha_1;\sigma_2,\alpha_2}(\textbf{k},\omega;\textbf{Q},\Omega) =-F_{\sigma_2,\alpha_2;\sigma_1,\alpha_1}(-\textbf{k},-\omega;\textbf{Q},\Omega)
\label{eq:bere} 
\end{equation}  
where $\textbf{k}$ and $\omega$ ($\textbf{Q}$ and $\Omega$) are the momentum and frequency conjugate to the relative (average) position and time coordinates. Eq.~(\ref{eq:bere}) encapsulates the generalization of the Berezinskii classification to systems which lack translation symmetry in space and time. We note that there is no requirement on the symmetry with respect to the average momenta or frequency. Thus if the dependence on $\textbf{Q}$ and $\Omega$ is suppressed this reduces to the standard Berezinskii classification.

With a framework now established to classify the symmetries of a driven superconductor we would like to study a concrete example of a driven system with pair amplitudes falling into one of the odd-$\omega$ classes. Heuristically, one might expect that subjecting a conventional superconductor to a time-dependent potential could lead to the creation of odd-$\omega$ pair amplitudes since the correlation functions will then develop a nontrivial dependence on the absolute time arguments through convolutions with the driving potential. This time-dependence will then manifest itself as additional pair amplitudes that must each fall into one of the 8 symmetry classes outlined in Table~\ref{table:classification} and, a priori, there is no reason to exclude the possible emergence of odd-$\omega$ terms. However, we note that if the driving potential acts trivially on orbital and spin degrees of freedom then, in the case where the original superconductor is a spin singlet, and even in both parity and time ($\mathcal{S}=-1$, $\mathcal{T}=1$, $\mathcal{O}=1$, $\mathcal{P}=1$), the only possible terms that are odd-$\omega$ would be in the class defined by ($\mathcal{S}=-1$, $\mathcal{T}=-1$, $\mathcal{O}=1$, $\mathcal{P}=-1$). Therefore, we expect that the driving potential should also depend on position to generate odd-parity amplitudes. Alternatively, if the drive were allowed to act nontrivially on spin or orbital degrees of freedom it should be possible to generate odd-$\omega$ terms in the class defined by ($\mathcal{S}=1$, $\mathcal{T}=-1$, $\mathcal{O}=1$, $\mathcal{P}=1$) or ($\mathcal{S}=-1$, $\mathcal{T}=-1$, $\mathcal{O}=-1$, $\mathcal{P}=1$), respectively. We will now demonstrate this transmutation of symmetries explicitly for the case of a conventional superconductor driven by a spatially nonuniform AC electric field.    
\begin{figure}
 \begin{center}
  \centering
  \includegraphics[width=0.5\textwidth]{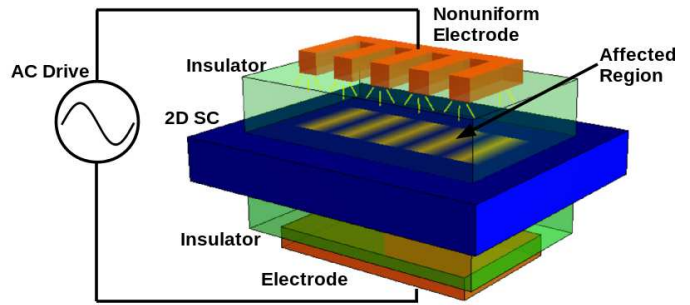}
  \caption{(color online) Schematic of a driven superconducting system with a 2D superconducting region described by Eq.~(\ref{eq:Ham}) lying between two insulating slabs each capped by a conducting electrode configured in such a way as to generate a spatially nonuniform electric field. The AC voltage acts as a time-dependent drive. Such a device could be realized by sandwiching a thin film superconductor, like Pb\cite{zhang2005band,eom2006persistent,brun2009reduction,zhang2010superconductivity}, between two insulating wafers.  
         }
  \label{fig:schematic}
 \end{center}
\end{figure}

\subsection{Model for Driven Superconductor}

The schematic shown in Fig.~\ref{fig:schematic} represents a conceptually simple method for coupling a two-dimensional (2D) $s$-wave singlet superconductor to a time- and position-dependent drive. This system could be realized using a thin film superconductor, for example Pb\cite{zhang2005band,eom2006persistent,brun2009reduction,zhang2010superconductivity}, sandwiched between two insulators and the drive could be implemented by biasing the sample with a spatially nonuniform AC voltage. As we will show, in such a device we expect the Cooper pairs in the 2D superconductor to develop odd-$\omega$ correlations. The superconducting region in Fig.~\ref{fig:schematic} is described by the Hamiltonian $H=H_0+H_{i}+H_t$ where:
\begin{equation}
\begin{aligned}
H_0 &= \sum_{\textbf{r},\textbf{r}',\sigma}(\gamma_{\textbf{r}\textbf{r}'}-\mu \delta_{\textbf{r}\textbf{r}'} ) \psi^\dagger_{\sigma,\textbf{r}} \psi_{\sigma,\textbf{r}'} \\
H_i&= \sum_{\textbf{r}}\lambda \psi^\dagger_{\uparrow,\textbf{r}}\psi^\dagger_{\downarrow,\textbf{r}}\psi_{\downarrow,\textbf{r}} \psi_{\uparrow,\textbf{r}} \\
H_t &=\sum_{\textbf{r},\sigma} U_{\textbf{r},t}\psi^\dagger_{\sigma,\textbf{r}} \psi_{\sigma,\textbf{r}}
\end{aligned}
\label{eq:Ham}
\end{equation}
where $\psi^\dagger_{\sigma,\textbf{r}}$ ($\psi_{\sigma,\textbf{r}}$) creates (annihilates) a fermionic quasiparticle state with spin $\sigma$ on site $\textbf{r}$, $\gamma_{\textbf{r}\textbf{r}'}$ is the hopping integral describing the tight-binding model of the normal state quasiparticles with chemical potential $\mu$, $\lambda$ describes an attractive interaction between quasiparticles, and $U_{\textbf{r},t}$ is a time- and position-dependent potential modeling the effect of the AC voltage.

To study the case of a time-dependent drive, $U_{\textbf{r},t}$, we employ non-equilibrium Green's functions\cite{rammer2007quantum,stefanucci2013nonequilibrium}. In this formalism the Green's function can be represented as an 8$\times$8 matrix written in a combined Keldysh$\times$particle-hole$\times$spin space:
\begin{equation}
\hat{\mathcal{G}}(x_1;x_2)=\left(
\begin{array}{cc}
\hat{\mathcal{G}}^{R}(x_1;x_2) & \hat{\mathcal{G}}^{K}(x_1;x_2) \\
0 & \hat{\mathcal{G}}^{A}(x_1;x_2) 
\end{array} \right)
\label{eq:g_keldysh}
\end{equation}
where we adopt the notation $x_i\equiv(\textbf{r}_i,t_i)$ where $\textbf{r}_i$ is a position in the 2D superconductor and $t_i$ is a time defined on the interval $(-\infty,\infty)$, and where each component, $\hat{\mathcal{G}}^{\alpha}(x_1;x_2)$, is a matrix in particle-hole$\times$spin space:
\begin{equation}
\hat{\mathcal{G}}^{\alpha}(x_1;x_2)=\left(
\begin{array}{cc}
\hat{G}^{\alpha}(x_1;x_2) & \hat{F}^{\alpha}(x_1;x_2) \\
\hat{\overline{F}}^{\alpha}(x_1;x_2) & \hat{\overline{G}}^{\alpha}(x_1;x_2)  
\end{array} \right)
\label{eq:g_nambu}
\end{equation}
where the ``hat" indicates that each block is a 2$\times$2 matrix in spin space. The retarded and Keldysh entries are defined as:
\begin{equation}
\begin{aligned}
G^{R}_{\sigma_1 \sigma_2}(x_1;x_2) &= -i\theta(t_1-t_2)\langle \{\psi_{\sigma_1,\textbf{r}_1}(t_1),\psi^\dagger_{\sigma_2,\textbf{r}_2}(t_2) \}\rangle \\
F^{R}_{\sigma_1 \sigma_2}(x_1;x_2) &= -i\theta(t_1-t_2)\langle \{\psi_{\sigma_1,\textbf{r}_1}(t_1),\psi_{\sigma_2,\textbf{r}_2}(t_2) \}\rangle \\
\overline{G}^{R}_{\sigma_1 \sigma_2}(x_1;x_2) &= -i\theta(t_1-t_2)\langle \{\psi^\dagger_{\sigma_1,\textbf{r}_1}(t_1),\psi_{\sigma_2,\textbf{r}_2}(t_2) \}\rangle \\
\overline{F}^{R}_{\sigma_1 \sigma_2}(x_1;x_2) &= -i\theta(t_1-t_2)\langle \{\psi^\dagger_{\sigma_1,\textbf{r}_1}(t_1),\psi^\dagger_{\sigma_2,\textbf{r}_2}(t_2) \}\rangle \\
G^{K}_{\sigma_1 \sigma_2}(x_1;x_2) &= -i\langle [\psi_{\sigma_1,\textbf{r}_1}(t_1),\psi^\dagger_{\sigma_2,\textbf{r}_2}(t_2) ]\rangle \\
F^{K}_{\sigma_1 \sigma_2}(x_1;x_2) &= -i\langle [\psi_{\sigma_1,\textbf{r}_1}(t_1),\psi_{\sigma_2,\textbf{r}_2}(t_2) ]\rangle \\
\overline{G}^{K}_{\sigma_1 \sigma_2}(x_1;x_2) &= -i\langle [\psi^\dagger_{\sigma_1,\textbf{r}_1}(t_1),\psi_{\sigma_2,\textbf{r}_2}(t_2) ]\rangle \\
\overline{F}^{K}_{\sigma_1 \sigma_2}(x_1;x_2) &= -i\langle [\psi^\dagger_{\sigma_1,\textbf{r}_1}(t_1),\psi^\dagger_{\sigma_2,\textbf{r}_2}(t_2) ]\rangle
\end{aligned}
\label{eq:g_retkel}
\end{equation}
while the advanced block is related to the retarded block by $\hat{\mathcal{G}}^{A}(x_1;x_2)=\hat{\mathcal{G}}^{R}(x_2;x_1)^\dagger$.

To describe the superconducting state we decouple the four-fermion term in Eq.(\ref{eq:Ham}) using mean field theory. In this case we find that $\lambda\psi^\dagger_{\uparrow,\textbf{r}}\psi^\dagger_{\downarrow,\textbf{r}}\psi_{\downarrow,\textbf{r}} \psi_{\uparrow,\textbf{r}}\approx\Delta(\textbf{r},t)\psi^\dagger_{\uparrow,\textbf{r}}\psi^\dagger_{\downarrow,\textbf{r}}+\text{h.c.}$ where $\Delta(\textbf{r},t)=\lambda\langle \psi_{\downarrow,\textbf{r}}(t)\psi_{\uparrow,\textbf{r}}(t)\rangle$. Written in terms of the Green's functions defined in Eqs (\ref{eq:g_keldysh}) and (\ref{eq:g_nambu}), and making the position- and time-dependence explicit, we have:
\begin{equation}
\Delta(\textbf{r},t)=i\frac{\lambda}{2}\left( F^{R}_{\downarrow\uparrow}(\textbf{r},t;\textbf{r},t) - F^{A}_{\downarrow\uparrow}(\textbf{r},t;\textbf{r},t) + F^{K}_{\downarrow\uparrow}(\textbf{r},t;\textbf{r},t) \right).
\label{eq:gap}
\end{equation}
 
With the mean field defined in Eq (\ref{eq:gap}) and using the Heisenberg equations of motion for the operators $\psi_{\sigma,\textbf{r}}(t)$, one can show that the equations of motion for the Green's functions in Eq (\ref{eq:g_keldysh}) are:
\begin{widetext}
\begin{equation}
\begin{aligned}
\hat{\mathcal{G}}^{-1}(x_1)\hat{\mathcal{G}}(x_1;x_2)&=\delta(x_1-x_2) \ \hat{\sigma}_0\otimes\hat{\rho}_0 \otimes\hat{\tau}_0 ; \\
\hat{\mathcal{G}}^{-1}(x_1) &= \left( \begin{array}{cc}
i\dfrac{d}{dt_1} - \left(-\dfrac{\nabla^2_{\textbf{r}_1}}{2m} -\mu-U_{\textbf{r}_1,t_1} \right) & -\Delta(\textbf{r}_1,t_1)i\hat{\sigma}_2 \\
-(\Delta(\textbf{r}_1,t_1)i\hat{\sigma}_2)^\dagger & i\dfrac{d}{dt_1} + \left(-\dfrac{\nabla^2_{\textbf{r}_1}}{2m} -\mu-U_{\textbf{r}_1,t_1} \right)\end{array}\right)\otimes \hat{\tau}_0
\label{eom}
\end{aligned}
\end{equation}
\end{widetext}
where $\hat{\sigma}_0$, $\hat{\rho}_0$, and $\hat{\tau}_0$ are the identity matrices in spin, particle-hole, and Keldysh space respectively, $\hat{\sigma}_{i\neq0}$ are the Pauli matrices in spin space, and we have approximated the hopping integral $\gamma_{\textbf{r},\textbf{r}'}$ by $\nabla^2_{\textbf{r}_1}/2m$.

In the absence of a drive ($U_{\textbf{r},t}=0$) the system is in equilibrium and translation invariant, therefore $\Delta(\textbf{r},t)=\Delta_0$ is constant. In this unperturbed case one can Fourier transform Eq (\ref{eom}) and find:
\begin{widetext}
\begin{equation}
\begin{aligned}
\hat{\mathcal{G}}_0^R(\textbf{k},\omega)&= \lim_{\eta \rightarrow 0^+}\frac{1}{(\omega+i\eta)^2-\xi^2_{\textbf{k}}-|\Delta_0|^2}\left( \begin{array}{cc}
\left( \omega +i\eta + \xi_\textbf{k} \right)\hat{\sigma}_0 & \Delta_0 i\hat{\sigma}_2 \\
-\Delta_0^{*} i\hat{\sigma}_2 & \left( \omega +i\eta - \xi_\textbf{k} \right)\hat{\sigma}_0
\end{array} \right) \\
\hat{\mathcal{G}}_0^A(\textbf{k},\omega)&= \lim_{\eta \rightarrow 0^+}\frac{1}{(\omega-i\eta)^2-\xi^2_{\textbf{k}}-|\Delta_0|^2}\left( \begin{array}{cc}
\left( \omega -i\eta + \xi_\textbf{k} \right)\hat{\sigma}_0 & \Delta_0 i\hat{\sigma}_2 \\
-\Delta_0^{*} i\hat{\sigma}_2 & \left( \omega -i\eta - \xi_\textbf{k} \right)\hat{\sigma}_0
\end{array} \right) \\
\hat{\mathcal{G}}_0^K(\textbf{k},\omega)&=\frac{-i\pi\tanh\left(\frac{\omega\beta}{2}\right)}{\sqrt{\xi_{\textbf{k}}^2 + |\Delta_0|^2}} \left[ \delta\left(\omega-\sqrt{\xi_\textbf{k}^2 +|\Delta_0|^2}\right) - \delta\left(\omega+\sqrt{\xi_\textbf{k}^2 +|\Delta_0|^2}\right)\right] \left( \begin{array}{cc}
\left(\omega+\xi_\textbf{k}\right)\hat{\sigma}_0 & \Delta_0 i\hat{\sigma}_2 \\
-\Delta_0^{*} i\hat{\sigma}_2 & \left(\omega -\xi_\textbf{k} \right)\hat{\sigma}_0
\end{array} \right)
\end{aligned}
\label{eq:rak}
\end{equation}
\end{widetext}
where $\xi_\textbf{k}=\frac{\hbar^2}{2m}k^2-\mu$ and $\beta$ is the inverse temperature of the superconductor.

In the presence of a time-dependent drive, $U_{\textbf{r},t}$, the Green's functions in Eq (\ref{eom}) can be related to the solutions to the undriven problem by the Dyson equation:
\begin{equation}
\hat{\mathcal{G}}(x_1;x_2)=\hat{\mathcal{G}}_0(x_1;x_2) - \int dx \hat{\mathcal{G}}_0(x_1;x) \hat{U}_{x} \hat{\mathcal{G}}(x;x_2) \\
\label{eq:dyson}
\end{equation}
where $\int dx=\int dt \int d\textbf{r}$, $\hat{U}_{x}=U_{\textbf{r},t} \ \hat{\sigma}_0\otimes\hat{\rho}_3\otimes\hat{\tau}_0$, and $\hat{\rho}_{3}$ is the Pauli matrix in particle-hole space. 

For a strong drive, $U_{\textbf{r},t}$, Eq (\ref{eq:dyson}) must be solved with a self-consistent solution for $\Delta(\textbf{r},t)$ in which case Eq (\ref{eq:rak}) would no longer be valid. However, as shown in the appendix, the first order self-consistent corrections to $\Delta(\textbf{r},t)$ are given by:
\begin{equation}
\frac{\delta\Delta(\textbf{r},t)}{\Delta_0}=\frac{-\lambda N_0}{4}\frac{U_0}{\mu} e^{-i\textbf{Q}_0\cdot\textbf{r}-i\Omega_0 t} \left[2+\frac{\Omega_0}{2\mu}+\frac{\pi^2}{12}\overline{Q}_0^2\frac{\Delta_0^2}{\mu^2} \right] 
\label{eq:delta_delta}
\end{equation}    
where $N_0=m/2\pi\hbar^2$ is the density of states, $\overline{Q}_0$ is the driving wavevector times the coherence length $\xi=\hbar v_F/\pi\Delta_0$, $\Omega_0$ is the driving frequency, and $U_0$ is the amplitude of the drive. 

We note that in superconducting Pb thin films typical values for these parameters are: $\lambda N_0\approx 1$\cite{zhang2005band,zhang2010superconductivity}, $\mu\approx 1.2$eV\cite{zhang2005band}, $\Delta_0\approx 1.17$meV\cite{eom2006persistent,brun2009reduction}, and $\xi\approx 49$nm\cite{zhang2010superconductivity}. Thus, for $\Omega_0<\mu$ ($<290 \text{THz}$) and $Q_0<\xi^{-1}$ ($<2\times 10^{-2}\text{nm}^{-1}$), $U_0$ can be as large as 100 meV and still provide adequate suppression of $\delta\Delta(\textbf{r},t)$ relative to $\Delta_0$. These values are well within experimental feasibility and so we may neglect the self-consistent corrections to $\Delta(\textbf{r},t)$.

Since we are justfied in assuming $\Delta(\textbf{r},t)=\Delta_0$ the Green's function $\hat{\mathcal{G}}_0(x_1;x_2)$ appearing in Eq (\ref{eq:dyson}) is simply given by Eq (\ref{eq:rak}). By iterating Eq (\ref{eq:dyson}) it is straightforward to compute the full Green's function $\hat{\mathcal{G}}(x_1;x_2)$ in terms of the unperturbed Green's function $\hat{\mathcal{G}}_0(x_1;x_2)$ to aribitrary order in the drive $U_{\textbf{r},t}$. For $|U_{\textbf{r},t}|<<\mu$ we truncate the series to linear order in $U_{\textbf{r},t}$, and thus:
\begin{widetext}
\begin{equation}
\hat{\mathcal{G}}(\textbf{q},\omega;\textbf{Q},\Omega)\approx\delta(\textbf{Q})\delta(\Omega)\hat{\mathcal{G}}_0(\textbf{q};\omega) - \hat{\mathcal{G}}_0(\textbf{q}+\tfrac{\textbf{Q}}{2};\omega+\tfrac{\Omega}{2}) \hat{U}_{\textbf{Q},\Omega} \hat{\mathcal{G}}_0(\textbf{q}-\tfrac{\textbf{Q}}{2};\omega-\tfrac{\Omega}{2}) \\
\label{eq:dyson_linear}
\end{equation}
\end{widetext} 
where $\textbf{q}$ and $\omega$ are the momentum and frequency conjugate to the relative position, $\textbf{r}_1-\textbf{r}_2$, and time, $t_1-t_2$, $\textbf{Q}$ and $\Omega$ are the momentum and frequency conjugate to the average position, $(\textbf{r}_1+\textbf{r}_2)/2$, and time, $(t_1+t_2)/2$, $\hat{U}_{\textbf{k},\omega}=\int d\textbf{r} dt e^{it\omega+i\textbf{r}\cdot\textbf{k}} \hat{U}_{\textbf{r},t}$, and $\hat{\mathcal{G}}_0(\textbf{q};\omega)$ is given by Eq (\ref{eq:rak}).

Focusing on the retarded components of $\hat{\mathcal{G}}(\textbf{q},\omega;\textbf{Q},\Omega)$ in Eq (\ref{eq:dyson_linear}) we find that the corrections to linear order in the drive are given by:
\begin{widetext}
\begin{equation}
\begin{aligned}
\delta\hat{G}^{R}(\textbf{q},\omega;\textbf{Q},\Omega) &= U_{\textbf{Q},\Omega}  \left[ \hat{G}^{R}_0(\textbf{q}+\tfrac{\textbf{Q}}{2};\omega+\tfrac{\Omega}{2})\hat{G}^{R}_0(\textbf{q}-\tfrac{\textbf{Q}}{2};\omega-\tfrac{\Omega}{2}) - \hat{F}^{R}_0(\textbf{q}+\tfrac{\textbf{Q}}{2};\omega+\tfrac{\Omega}{2})\hat{\overline{F}}^{R}_0(\textbf{q}-\tfrac{\textbf{Q}}{2};\omega-\tfrac{\Omega}{2}) \right] \\
\delta\hat{F}^{R}(\textbf{q},\omega;\textbf{Q},\Omega) &= U_{\textbf{Q},\Omega}  \left[ \hat{G}^{R}_0(\textbf{q}+\tfrac{\textbf{Q}}{2};\omega+\tfrac{\Omega}{2})\hat{F}^{R}_0(\textbf{q}-\tfrac{\textbf{Q}}{2};\omega-\tfrac{\Omega}{2}) - \hat{F}^{R}_0(\textbf{q}+\tfrac{\textbf{Q}}{2};\omega+\tfrac{\Omega}{2})\hat{\overline{G}}_0^{R}(\textbf{q}-\tfrac{\textbf{Q}}{2};\omega-\tfrac{\Omega}{2}) \right] \\
\delta\hat{\overline{G}}^{R}(\textbf{q},\omega;\textbf{Q},\Omega) &= U_{\textbf{Q},\Omega}  \left[ \hat{\overline{F}}^{R}_0(\textbf{q}+\tfrac{\textbf{Q}}{2};\omega+\tfrac{\Omega}{2})\hat{F}^{R}_0(\textbf{q}-\tfrac{\textbf{Q}}{2};\omega-\tfrac{\Omega}{2}) - \hat{\overline{G}}^{R}_0(\textbf{q}+\tfrac{\textbf{Q}}{2};\omega+\tfrac{\Omega}{2})\hat{\overline{G}}^{R}_0(\textbf{q}-\tfrac{\textbf{Q}}{2};\omega-\tfrac{\Omega}{2})\right] \\
\delta\hat{\overline{F}}^{R}(\textbf{q},\omega;\textbf{Q},\Omega) &= U_{\textbf{Q},\Omega}  \left[ \hat{\overline{F}}^{R}_0(\textbf{q}+\tfrac{\textbf{Q}}{2};\omega+\tfrac{\Omega}{2})\hat{G}_0^{R}(\textbf{q}-\tfrac{\textbf{Q}}{2};\omega-\tfrac{\Omega}{2}) - \hat{\overline{G}}^{R}_0(\textbf{q}+\tfrac{\textbf{Q}}{2};\omega+\tfrac{\Omega}{2})\hat{\overline{F}}^{R}_0(\textbf{q}-\tfrac{\textbf{Q}}{2};\omega-\tfrac{\Omega}{2}) \right].
\end{aligned}
\label{eq:r_linear}
\end{equation}
\end{widetext}

\subsection{Odd-frequency Terms in the Anomalous Green's Function}

Inserting the expressions from Eq (\ref{eq:rak}) to the equation for the linear corrections to the anomalous Green's function in Eq (\ref{eq:r_linear}) we find:
\begin{equation}
\delta\hat{F}^{R}(\textbf{q},\omega;\textbf{Q},\Omega) = i\hat{\sigma}_2   \dfrac{U_{\textbf{Q},\Omega} \Delta_0\left(\Omega + \xi_{\textbf{q}+\frac{\textbf{Q}}{2}} + \xi_{\textbf{q}-\frac{\textbf{Q}}{2}}\right) }{D_{\textbf{q},\omega;\textbf{Q},\Omega}D_{\textbf{q},\omega;-\textbf{Q},-\Omega}} 
\label{eq:fr_linear}
\end{equation}
where 
\begin{equation}
D_{\textbf{q},\omega;\textbf{Q},\Omega}\equiv(\omega+\tfrac{\Omega}{2})^2-\xi_{\textbf{q}+\frac{\textbf{Q}}{2}}^2-\Delta_0^2.
\label{eq:denomenator}
\end{equation} 
By taking $\omega\rightarrow-\omega$ in Eq (\ref{eq:fr_linear}) we can see that this expression is neither even nor odd in $\omega$ and thus we have shown that the pair correlation function describing the driven condensate has both even- and odd- frequency terms.

It is straightforward to show that the odd-$\omega$ terms in $\delta\hat{F}^{R}(\textbf{q},\omega;\textbf{Q},\Omega)$ are given by:
\begin{equation}
\delta\hat{F}^{R}_{odd}(\textbf{q},\omega;\textbf{Q},\Omega) = i\hat{\sigma}_2 \ \mathcal{C}_{\textbf{q},\omega;\textbf{Q},\Omega} \ \omega\Omega\left( \xi_{\textbf{q}+\frac{\textbf{Q}}{2}} - \xi_{\textbf{q}-\frac{\textbf{Q}}{2}}\right)
\label{eq:odd_term}
\end{equation}
where we define the coefficient $\mathcal{C}_{\textbf{q},\omega;\textbf{Q},\Omega}= \frac{ U_{\textbf{Q},\Omega}\Delta_0\left(\Omega + \xi_{\textbf{q}+\frac{\textbf{Q}}{2}} + \xi_{\textbf{q}-\frac{\textbf{Q}}{2}}\right)}{D_{\textbf{q},\omega;\textbf{Q},\Omega}D_{\textbf{q},\omega;\textbf{Q},-\Omega}D_{\textbf{q},\omega;-\textbf{Q},\Omega} D_{\textbf{q},\omega;-\textbf{Q},-\Omega} } $ which is strictly even in $\omega$ and $\textbf{q}$. 

Written in this form Eq (\ref{eq:odd_term}) clearly has singlet spin structure and is manifestly odd in $\omega$. Additionally, by taking $\textbf{q}\rightarrow-\textbf{q}$ and noting that $\xi_{-\textbf{q}}=\xi_{\textbf{q}}$ it is clear that this term is odd in parity. Hence, starting from a superconductor with only spin singlet even-parity even-$\omega$ ($\mathcal{S}=-1$, $\mathcal{T}=1$, $\mathcal{O}=1$, $\mathcal{P}=1$) pairs we have induced spin singlet odd-parity odd-$\omega$ ($\mathcal{S}=-1$, $\mathcal{T}=-1$, $\mathcal{O}=1$, $\mathcal{P}=-1$) pair amplitudes. We note that if $|\textbf{Q}|\rightarrow 0$ or $\Omega\rightarrow 0$ the ($\mathcal{S}=-1$, $\mathcal{T}=-1$, $\mathcal{O}=1$, $\mathcal{P}=-1$) amplitudes will vanish which is consistent with our expectation that the drive must possess both time-dependence and spatial inhomogeneity to transmute even-parity even-$\omega$ pairs into odd-parity odd-$\omega$ pairs.  

Furthermore, we note that the odd-$\omega$ terms in $\delta\hat{F}^{R}(\textbf{q},\omega;\textbf{Q},\Omega)$ are a direct consequence of the denominator $D_{\textbf{q},\omega;\textbf{Q},\Omega}$ which acts to shift the spectrum in both frequency and momentum space. This denominator is shared by both $\delta\hat{F}^{R}(\textbf{q},\omega;\textbf{Q},\Omega)$ and $\delta\hat{G}^{R}(\textbf{q},\omega;\textbf{Q},\Omega)$ which is connected to observables. In section \ref{observables} we will explore measurable consequences of these shifted spectra.  

\subsection{Time-Dependence of $\delta\hat{F}^{R}_{odd}$}

Since $\delta\hat{F}^{R}_{odd}(\textbf{q},\omega;\textbf{Q},\Omega)$ in Eq (\ref{eq:odd_term}) depends on the center-of-mass frequency ($\Omega$) and momentum ($\textbf{Q}$) we can deduce that it must have some nontrivial dependence on absolute time and position. We will now analyze this time- and space-dependence by transforming Eq (\ref{eq:odd_term}) to the mixed representation using the definition:
\begin{equation}
\hat{\mathcal{G}}(\textbf{q},\omega;\textbf{R},T)\equiv \int \frac{d\Omega}{2\pi}\frac{d\textbf{Q}}{(2\pi)^2} \hat{\mathcal{G}}(\textbf{q},\omega;\textbf{Q},\Omega) e^{-i\textbf{R}\cdot\textbf{Q}-iT\Omega}
\label{eq:mixed}
\end{equation}    
together with a concrete expression for the drive:
\begin{equation}
U_{\textbf{r},t}=U_0\cos\left( \textbf{Q}_0\cdot\textbf{r} \right)\cos\left( \Omega_0 t \right).
\label{eq:drive}
\end{equation}

Using the expressions in Eqs (\ref{eq:odd_term})-(\ref{eq:drive}) we find that the odd-$\omega$ terms in $\delta\hat{F}^{R}(\textbf{q},\omega;\textbf{R},T)$ are given by:
\begin{widetext}
\begin{equation}
\delta\hat{F}^{R}_{odd}(\textbf{q},\omega;\textbf{R},T)= \omega \Delta_0i\hat{\sigma}_2 \frac{ U_0 \Omega_0  \sin(\textbf{Q}_0\cdot\textbf{R})  \left(  \xi_{\textbf{q}+\tfrac{\textbf{Q}_0}{2}} - \xi_{\textbf{q}-\tfrac{\textbf{Q}_0}{2}} \right) \left[ \sin(\Omega_0T)\left( \xi_{\textbf{q}+\tfrac{\textbf{Q}_0}{2}} + \xi_{\textbf{q}-\tfrac{\textbf{Q}_0}{2}} \right) + i\cos(\Omega_0T) \Omega_0 \right] }{D_{\textbf{q},\omega;\textbf{Q}_0,\Omega_0}D_{\textbf{q},\omega;\textbf{Q}_0,-\Omega_0}D_{\textbf{q},\omega;-\textbf{Q}_0,\Omega_0} D_{\textbf{q},\omega;-\textbf{Q}_0,-\Omega_0}} .
\label{eq:fr_mixed}
\end{equation}
\end{widetext}
Notice that $\delta\hat{F}^{R}_{odd}(\textbf{q},\omega;\textbf{R},T)$ possesses a finite real part which evolves in time, $T$, 90$^\circ$ out-of-phase with respect to the imaginary part of $\delta\hat{F}^{R}_{odd}(\textbf{q},\omega;\textbf{R},T)$ while both the real and imaginary parts possess the same spatial modulation. If we focus on the real part of Eq (\ref{eq:fr_mixed}) we see that it is exactly 90$^\circ$ out-of-phase with the drive and thus the odd-$\omega$ correlations are largest at the times and positions where the drive vanishes.

\section{Observable Features}
\label{observables}

We will now investigate the relationship between the dynamically-induced odd-$\omega$ pairing and two measureable quantities: the spectral function and density of states.

\subsection{Spectral Function}

In general, the spectral function can be related to the retarded components of the Green's function by:
\begin{equation}
A_{\textbf{k},\omega}(\textbf{R},T)=-\frac{1}{\pi}\text{Im}\text{Tr} \hat{\mathcal{G}}^{R}(\textbf{k},\omega;\textbf{R},T) 
\label{eq:akw}
\end{equation}
where $\textbf{R}$ and $T$ are the average position and time. This quantity can be measured using photoemission spectroscopy\cite{damascelli2003angle}. 

\begin{figure}
 \begin{center}
  \centering
  \includegraphics[width=0.5\textwidth]{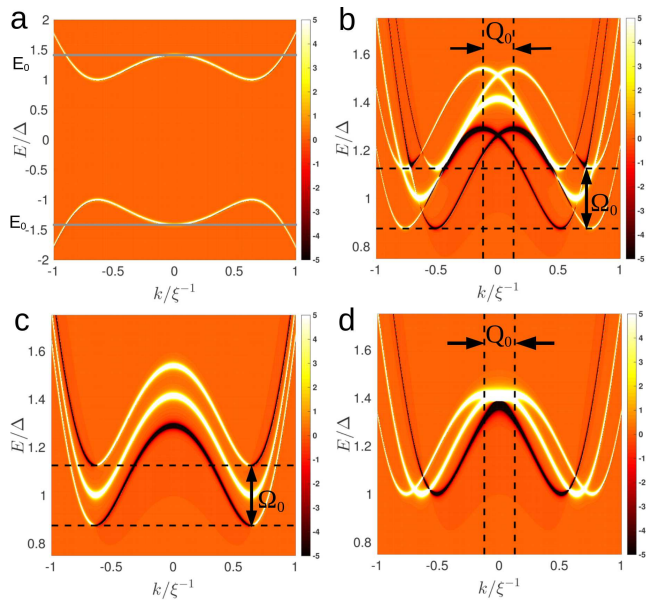}
  \caption{(color online) Comparison of the spectral function, $A_{\textbf{k},\omega}(\textbf{R},T)$, for both an undriven and driven 2D superconductor described by the Hamiltonian in Eq.~(\ref{eq:Ham}). All energies are presented in units of the undriven order parameter, $\Delta_0$, all wavevectors are presented in units of the inverse coherence length $\xi^{-1}=\pi\Delta_0/\hbar v_F$ and plotted along the $k_x$-direction. (a) Spectral function for an undriven $s$-wave superconductor using Eq (\ref{eq:a0}). Horizontal lines at $\pm E_{0}=\pm\sqrt{\mu^2 + \Delta_0^2}$ denote the energies associated with notable extrema of the ``sombrero" shape. (b), (c), \& (d) Spectral functions for a driven $s$-wave superconductor calculated using Eq (\ref{eq:akwrt}) for $\textbf{R}=0$, $T=0$, with $U=0.25$, but for different pairs of $\Omega_0$ and $\textbf{Q}_0$: (b) $\Omega_0=0.25$, $\textbf{Q}_0=(0.25,0)$; (c) $\Omega_0=0.25$, $\textbf{Q}_0=0$; and (d) $\Omega_0=0$, $\textbf{Q}_0=(0.25,0)$. In each case vertical and horizontal lines denote the shifted spectra arising at $\pm\textbf{Q}_0/2$ and $\pm\Omega_0/2$.  
         }
  \label{fig:spec}
 \end{center}
\end{figure}

In the absence of an external drive, $A_{\textbf{k},\omega}(\textbf{R},T)$ for a mean-field superconductor is given by:
\begin{equation}
A^{(0)}_{\textbf{k},\omega}=2\left[ \delta\left( \omega - E_\textbf{k} \right) + \delta\left( \omega + E_\textbf{k} \right) \right]
\label{eq:a0}
\end{equation}
where $E_\textbf{k}=\sqrt{\xi_{\textbf{k}}^2 + \Delta_0^2}$. As we can see from Fig~\ref{fig:spec} (a), this spectral function exhibits two bands of quasiparticle states separated by a gap of $2\Delta_0$. These bands are each defined by the standard ``sombrero" shape with extrema appearing at $k=0$, $E_0=\pm\sqrt{\mu^2+\Delta_0^2}$ and $k=\sqrt{2m\mu}/\hbar$, $E_{k_F}=\pm\Delta_0$.  

We will account for the presence of an external drive by using the expressions for the undriven Green's functions in Eq (\ref{eq:rak}) together with the linear order corrections to the retarded Green's functions in Eq (\ref{eq:r_linear}) to obtain the linear order corrections to $A_{\textbf{k},\omega}(\textbf{R},T)$:
\begin{widetext}
\begin{equation}
\begin{aligned}
A_{\textbf{k},\omega}(\textbf{R},T) &=A^{(0)}_{\textbf{k},\omega} + \cos(\textbf{R}\cdot\textbf{Q}_0+T\Omega_0)\delta A^{(1a)}_{\textbf{k},\omega} + \cos(\textbf{R}\cdot\textbf{Q}_0-T\Omega_0)\delta A^{(1b)}_{\textbf{k},\omega}, \\
\delta A^{(1a)}_{\textbf{k},\omega} &=\left[ f_{\textbf{k},\omega}(\textbf{Q}_0,\Omega_0) \delta\left( \omega + \frac{\Omega_0}{2} + E_{\textbf{k}+\tfrac{\textbf{Q}_0}{2}}\right) + f_{\textbf{k},\omega}(\textbf{Q}_0,-\Omega_0) \delta\left( \omega + \frac{\Omega_0}{2} - E_{\textbf{k}+\tfrac{\textbf{Q}_0}{2}}\right) \right. \\
&+\left. f_{\textbf{k},\omega}(-\textbf{Q}_0,\Omega_0) \delta\left( \omega - \frac{\Omega_0}{2} - E_{\textbf{k}-\tfrac{\textbf{Q}_0}{2}}\right) + f_{\textbf{k},\omega}(-\textbf{Q}_0,-\Omega_0) \delta\left( \omega - \frac{\Omega_0}{2} + E_{\textbf{k}-\tfrac{\textbf{Q}_0}{2}}\right) \right], \\
\delta A^{(1b)}_{\textbf{k},\omega} &=\left[ f_{\textbf{k},\omega}(\textbf{Q}_0,\Omega_0) \delta\left( \omega - \frac{\Omega_0}{2} - E_{\textbf{k}+\tfrac{\textbf{Q}_0}{2}}\right) + f_{\textbf{k},\omega}(\textbf{Q}_0,-\Omega_0) \delta\left( \omega - \frac{\Omega_0}{2} + E_{\textbf{k}+\tfrac{\textbf{Q}_0}{2}}\right) \right. \\
&+\left. f_{\textbf{k},\omega}(-\textbf{Q}_0,\Omega_0) \delta\left( \omega + \frac{\Omega_0}{2} + E_{\textbf{k}-\tfrac{\textbf{Q}_0}{2}}\right) + f_{\textbf{k},\omega}(-\textbf{Q}_0,-\Omega_0) \delta\left( \omega + \frac{\Omega_0}{2} - E_{\textbf{k}-\tfrac{\textbf{Q}_0}{2}}\right) \right]
\end{aligned}
\label{eq:akwrt}
\end{equation}
\end{widetext} 
where we have assumed a drive of the form shown in Eq (\ref{eq:drive}) and, for convenience, we have defined the function: 
$$f_{\textbf{k},\omega}(\textbf{Q},\Omega)=U_0\frac{\xi_{\textbf{k}-\tfrac{\textbf{Q}}{2}} + \xi_{\textbf{k}+\tfrac{\textbf{Q}}{2}}\left(1+\dfrac{\Omega}{E_{\textbf{k}+\tfrac{\textbf{Q}}{2}}}\right) }{E^2_{\textbf{k}-\tfrac{\textbf{Q}}{2}}-E^2_{\textbf{k}+\tfrac{\textbf{Q}}{2}}\left(1+\dfrac{\Omega}{E_{\textbf{k}+\tfrac{\textbf{Q}}{2}}}\right)^2}.$$

In Fig~\ref{fig:spec}(b) we plot Eq (\ref{eq:akwrt}) for $\textbf{R}=0$ and $T=0$ and find that the main effect of the driving potential is the generation of duplicate spectra shifted in momentum and frequency by $\pm\textbf{Q}_0/2$ and $\pm\Omega_0/2$ respectively. From the form of Eq (\ref{eq:akwrt}) this is clear because the delta functions apearing in $\delta A^{(1a)}_{\textbf{k},\omega}$ and $\delta A^{(1b)}_{\textbf{k},\omega}$ determine which values of $\textbf{k}$ and $\omega$ obtain nonzero spectral weight. Physically this means that the shifted spectral features are due to the driving field imparting energy $\pm\Omega_0/2$ and momentum $\pm\textbf{Q}_0/2$ to the undriven quasiparticles. These shifted spectral features are in fact just the first harmonics in an infinite series emerging from higher order powers in the drive. 

It should be noted that these shifted spectral features can be traced back to the denominator from Eq (\ref{eq:denomenator}). Therefore, we can see a relation between the emergence of these harmonics and the odd-frequency terms in the anomalous Green's function. However, it is only for drives which possess nonzero $\textbf{Q}_0$ and $\Omega_0$, like the case shown in Fig~\ref{fig:spec} (b), that the generation of duplicate spectra coincides with the emergence of odd-$\omega$ terms in the anomalous Green's function.  

In Fig.~\ref{fig:spec} (c) we show $A_{\textbf{k},\omega}(\textbf{R},T)$ for the case of finite driving frequency and zero wavevector, $\Omega_0\neq0$ and $\textbf{Q}_0=0$, which corresponds to the case of driving the superconductor with a light source. In Fig.~\ref{fig:spec} (d) we plot the opposite limit of zero driving frequency and finite wavevector, $\Omega_0=0$ and $\textbf{Q}_0\neq0$, which corresponds to subjecting the superconductor to a static spatially nonuniform electric field. Notice that in both cases additional spectra are generated as expected from Eq (\ref{eq:akwrt}).   

Turning our attention to the case of $\textbf{R},T\neq0$, we can see from Eq (\ref{eq:akwrt}) that $A_{\textbf{k},\omega}(\textbf{R},T)$ possess periodic modulation in both $\textbf{R}$ and $T$. In the special case where $\textbf{R}\cdot\textbf{Q}_0+T\Omega_0=\textbf{R}\cdot\textbf{Q}_0-T\Omega_0=(2n+1)\pi/2$ we can see that the linear order corrections to the spectral function vanish and only the undriven spectrum remains. However, for all other values of $\textbf{R}$ and $T$ the additional spectra remain. Furthermore, in the case where $\textbf{R}\cdot\textbf{Q}_0+T\Omega_0=\textbf{R}\cdot\textbf{Q}_0-T\Omega_0=2n\pi$ we get exactly the results appearing in Fig~\ref{fig:spec}. For other values of $\textbf{R}$ and $T$ the relative weight and signs of the different shifted bands will change; however, the delta functions in Eq (\ref{eq:akwrt}) ensure that these features will be located at the same values of $\textbf{k}$ and $\omega$ as the features shown in Fig~\ref{fig:spec}. 

\subsection{Density of States}

Another important observable quantity is the electronic local density of states (LDOS), which can be calculated from the spectral function using
\begin{equation}
\mathcal{N}(\omega,\textbf{R},T)= \int \frac{d\textbf{q}}{(2\pi)^2} A_{\textbf{q},\omega}(\textbf{R},T)
\label{eq:dos}
\end{equation}
and can be measured by scanning tunneling microscopy\cite{hofer2003theories,tersoff1983theory}. 

\begin{figure}
 \begin{center}
  \centering
  \includegraphics[width=0.5\textwidth]{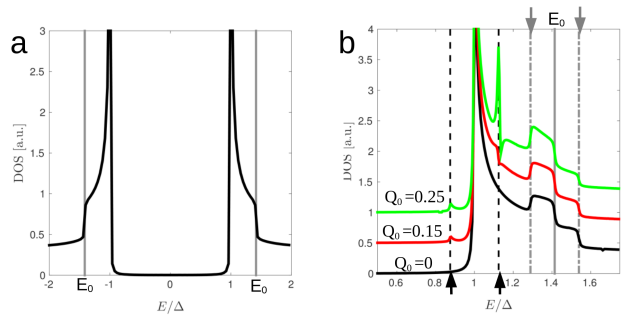}
  \caption{(color online) LDOS, $\mathcal{N}(\omega,\textbf{R}=0,T=0)$, for both the case of an undriven and driven 2D superconductor described by the Hamiltonian in Eq.~(\ref{eq:Ham}) with a driving potential of the form $U_{\textbf{r},t}=U\cos(\textbf{r}\cdot\textbf{Q}_0)\cos(t\Omega_0)$. (a) LDOS for undriven $s$-wave superconductor with vertical lines denoting the shoulders associated with the energies marked in Fig~\ref{fig:spec}(a). (b) LDOS for driven $s$-wave superconductor with $U=0.25$ and $\Omega_0=0.25$ evaluated for three different driving wavevectors $\textbf{Q}_0=(Q_0,0)$. Vertical lines mark the positions of: shoulder at $E_{0}$ (solid grey) and satelite features appearing at $E_0\pm\Omega_0/2$(dashed-dot grey), along with the satelite features associated with the coherence peak at $\Delta_0\pm\Omega_0/2$ (dashed black). Driven LDOS plots separated for clarity, in each case the LDOS vanishes within the gap.        
         }
  \label{fig:dos}
 \end{center}
\end{figure}

In Fig.~\ref{fig:dos} (a) we plot the LDOS for an undriven superconductor using Eqs (\ref{eq:a0}) and (\ref{eq:dos}). We can see that the undriven LDOS exhibits the trends of an $s$-wave superconductor, two coherence peaks separated by a gap of $2\Delta_0$. Notice the two shoulders at $\pm E_0$ just beyond the coherence peaks associated with the extrema at the center of the ``sombrero" shape in the spectra. 

In Fig.~\ref{fig:dos} (b) we show the driven LDOS at $\textbf{R}=0$, $T=0$, calculated using Eqs (\ref{eq:akwrt}) and (\ref{eq:dos}), for a single driving frequency of $\Omega_0=\Delta_0/4$ and overlay the plots for three different values of the driving wavevector $|\textbf{Q}_0|=Q_0$. The plots for different wavevectors have been shifted to make a clearer comparison, in each case the LDOS vanishes at the center of the gap. Two notable features emerge in the driven LDOS: (i) pairs of satelite peaks appear around the coherence peaks at $\pm\Delta_0\pm\Omega_0/2$; and (ii) pairs of singular points appear at $E_0\pm\Omega_0/2$. By comparing the features at different values of $Q_0$ we can see that the additional features at $E_0\pm\Omega_0/2$ hold for all $Q_0$ while the additional peak structures near the coherence peaks possess a strong dependence on $Q_0$. 

We note that these satelite coherence peaks arise only for nonzero $Q_0$ and $\Omega_0$, identically to the odd-$\omega$ terms in the anomalous Green's function. Furthermore, by comparing the plots in Figs~\ref{fig:spec} and \ref{fig:dos} we can see that the origin of these satelite coherence peaks appears to be from the positive spectral features shifted by $\textbf{Q}_0/2$ to the outside of the original spectra and then up and down in energy by $\Omega_0/2$. While we expect the satelite features near $E_0$ to appear in any driven superconductor regardless of the wavevector, the peaks appearing at $\Delta_0\pm\Omega_0/2$ seem to emerge only within the parameter regime in which the odd-$\omega$ terms are expected to develop in the pairing amplitude. This connection makes sense when we consider the fact that both the shifted spectral features and the odd-$\omega$ pairing terms have a common origin in the denominator of the retarded Green's function. Therefore, the shifted peaks in this system offer a possible means to infer the presence of the odd-$\omega$ pair amplitudes.

\section{Conclusions}

Motivated by the intrinsically dynamical nature of odd-frequency pairing we have generalized Berezinskii's classification of superconducting pairing symmetries to driven systems. Subsequently, we argued that under certain conditions odd-frequency pairing should arise as a consequence of a time-dependent driving potential and investigated this effect explicitly for a particular model of a conventional $s$-wave superconductor coupled to a time-dependent and spatially nonuniform electric field. Computing the corrections to the superconducting correlation functions in this system using linear response we found dynamically-induced odd-frequency terms appearing in the anomalous Green's function. Based on these results we can conclude that in order to generate odd-frequency pair amplitudes at this order using a spin-independent drive the external potential must be both time-dependent and spatially nonuniform. Furthermore, we demonstrated that these odd-frequency terms are related to features appearing in both the density of states and the spectral function, offering an experimental signature of their presence.

Acknowledgements: We wish to thank Annica Black-Schaffer, Matthias Eschrig, Mats Horsdal, Yaron Kedem, Sergey Pershoguba, and Enrico Rossi for useful discussions. This work was supported by US DOE BES E304 VR (AVB) and the European Research Council (ERC) DM-321031 (CT).

\appendix

\section{Self-consistent Corrections to the Superconducting Gap}

The results in this manuscript were obtained by expanding Eq (\ref{eq:dyson}) to linear order in the drive and evaluating this expression using the unperturbed Green's functions from Eq (\ref{eq:rak}). The unperturbed Green's functions, $\hat{\mathcal{G}}_0(\textbf{k},\omega)$, are obtained by Fourier transforming $\hat{\mathcal{G}}_0(x_1;x_2)$ which satisfies:
\begin{widetext}
\begin{equation}
\begin{aligned}
\hat{\mathcal{G}}_0^{-1}(x_1)\hat{\mathcal{G}}_0(x_1;x_2)&=\delta(x_1-x_2) \ \hat{\sigma}_0\otimes\hat{\rho}_0 \otimes\hat{\tau}_0 ; \\
\hat{\mathcal{G}}_0^{-1}(x_1) &= \left( \begin{array}{cc}
i\dfrac{d}{dt_1} - \left(-\dfrac{\nabla^2_{\textbf{r}_1}}{2m} -\mu \right) & -\Delta(\textbf{r}_1,t_1)i\hat{\sigma}_2 \\
-(\Delta(\textbf{r}_1,t_1)i\hat{\sigma}_2)^\dagger & i\dfrac{d}{dt_1} + \left(-\dfrac{\nabla^2_{\textbf{r}_1}}{2m} -\mu \right)\end{array}\right)\otimes \hat{\tau}_0
\end{aligned}
\label{eq:g0}
\end{equation}
\end{widetext}
where $\hat{\sigma}_0$, $\hat{\rho}_0$, and $\hat{\tau}_0$ are the identity matrices in spin, particle-hole, and Keldysh space respectively. 
In general we must solve Eq (\ref{eq:g0}) with a gap defined as:
\begin{equation}
\Delta(\textbf{r},t)=i\frac{\lambda}{2}\left( F^{R}_{\downarrow\uparrow}(\textbf{r},t;\textbf{r},t) - F^{A}_{\downarrow\uparrow}(\textbf{r},t;\textbf{r},t) + F^{K}_{\downarrow\uparrow}(\textbf{r},t;\textbf{r},t) \right)
\label{app:gap}
\end{equation}
where $F^{R}_{\downarrow\uparrow}(\textbf{r},t;\textbf{r},t)$,$F^{A}_{\downarrow\uparrow}(\textbf{r},t;\textbf{r},t)$,$F^{K}_{\downarrow\uparrow}(\textbf{r},t;\textbf{r},t)$ are solutions to the nonequilibrium problem. However the solutions to Eq (\ref{eq:g0}) presented in Eq (\ref{eq:rak}) were obtained assuming a constant gap $\Delta(\textbf{r},t)=\Delta_0$. First order deviations of $\Delta(\textbf{r},t)$ from this constant value are given by Eq (\ref{eq:delta_delta}) in the text. As discussed in the text, these deviations are negligible for a range of experimentally feasible drive potentials. In this appendix we will provide details about the calculation leading to Eq (\ref{eq:delta_delta}).   

From Eqs (\ref{app:gap}) and (\ref{eq:dyson_linear}) it can be shown that, to leading order in the drive, the gap can be written as
\begin{equation}
\Delta(\textbf{r},t)=\Delta_0 + \delta\Delta(\textbf{r},t) 
\end{equation}
where 
\begin{widetext}
\begin{equation}
\begin{aligned}
\delta\Delta(\textbf{r},t)&=-\frac{i\lambda}{2}\int\frac{d\textbf{q}}{(2\pi)^2}\frac{d\textbf{Q}}{(2\pi)^2}\frac{d\omega}{2\pi}\frac{d\Omega}{2\pi}e^{-i\textbf{Q}\cdot\textbf{r}-i\Omega t}\left[ \delta F^{R}(\textbf{q},\omega;\textbf{Q},\Omega)-\delta F^{A}(\textbf{q},\omega;\textbf{Q},\Omega)+\delta F^{K}(\textbf{q},\omega;\textbf{Q},\Omega) \right] 
\end{aligned}
\label{eq:gap_linear}
\end{equation}
where 
\begin{equation}
\begin{aligned}
\delta F^{R}(\textbf{q},\omega;\textbf{Q},\Omega)&=U_{\textbf{Q},\Omega}\left[G_0^{R}(\textbf{q}+\tfrac{\textbf{Q}}{2},\omega+\tfrac{\Omega}{2})F_0^{R}(\textbf{q}-\tfrac{\textbf{Q}}{2},\omega-\tfrac{\Omega}{2})-F_0^{R}(\textbf{q}+\tfrac{\textbf{Q}}{2},\omega+\tfrac{\Omega}{2})\overline{G}_0^{R}(\textbf{q}-\tfrac{\textbf{Q}}{2},\omega-\tfrac{\Omega}{2})\right] \\
\delta F^{A}(\textbf{q},\omega;\textbf{Q},\Omega)&=U_{\textbf{Q},\Omega}\left[G_0^{A}(\textbf{q}+\tfrac{\textbf{Q}}{2},\omega+\tfrac{\Omega}{2})F_0^{A}(\textbf{q}-\tfrac{\textbf{Q}}{2},\omega-\tfrac{\Omega}{2})-F_0^{A}(\textbf{q}+\tfrac{\textbf{Q}}{2},\omega+\tfrac{\Omega}{2})\overline{G}_0^{A}(\textbf{q}-\tfrac{\textbf{Q}}{2},\omega-\tfrac{\Omega}{2})\right] \\
\delta F^{K}(\textbf{q},\omega;\textbf{Q},\Omega)&=U_{\textbf{Q},\Omega}\left[G_0^{R}(\textbf{q}+\tfrac{\textbf{Q}}{2},\omega+\tfrac{\Omega}{2})F_0^{K}(\textbf{q}-\tfrac{\textbf{Q}}{2},\omega-\tfrac{\Omega}{2})-F_0^{R}(\textbf{q}+\tfrac{\textbf{Q}}{2},\omega+\tfrac{\Omega}{2})\overline{G}_0^{K}(\textbf{q}-\tfrac{\textbf{Q}}{2},\omega-\tfrac{\Omega}{2})\right. \\
&+\left.G_0^{K}(\textbf{q}+\tfrac{\textbf{Q}}{2},\omega+\tfrac{\Omega}{2})F_0^{A}(\textbf{q}-\tfrac{\textbf{Q}}{2},\omega-\tfrac{\Omega}{2})-F_0^{K}(\textbf{q}+\tfrac{\textbf{Q}}{2},\omega+\tfrac{\Omega}{2})\overline{G}_0^{A}(\textbf{q}-\tfrac{\textbf{Q}}{2},\omega-\tfrac{\Omega}{2})\right]
\end{aligned}
\label{eq:deltaF}
\end{equation}
where
\begin{equation}
\begin{aligned}
F^{K}_0(\textbf{k},\omega)&=\dfrac{i\pi \tanh\left(\frac{\beta \omega}{2} \right) \Delta_0}{\sqrt{\xi_\textbf{k}^2 +\Delta_0^2}} \left[\delta\left(\omega -\sqrt{\xi_\textbf{k}^2 +\Delta_0^2} \right) - \delta\left(\omega +\sqrt{\xi_\textbf{k}^2 +\Delta_0^2} \right)\right] \\
G^{K}_0(\textbf{k},\omega)&=\dfrac{-i\pi \tanh\left(\frac{\beta \omega}{2} \right)}{\sqrt{\xi_\textbf{k}^2 +\Delta_0^2}} \left[\left(\sqrt{\xi_\textbf{k}^2 +\Delta_0^2}+\xi_\textbf{k}\right)\delta\left(\omega -\sqrt{\xi_\textbf{k}^2 +\Delta_0^2} \right) + \left(\sqrt{\xi_\textbf{k}^2 +\Delta_0^2}-\xi_\textbf{k}\right)\delta\left(\omega +\sqrt{\xi_\textbf{k}^2 +\Delta_0^2} \right)\right] \\
\overline{G}^{K}_0(\textbf{k},\omega)&=\dfrac{-i\pi \tanh\left(\frac{\beta \omega}{2} \right)}{\sqrt{\xi_\textbf{k}^2 +\Delta_0^2}} \left[\left(\sqrt{\xi_\textbf{k}^2 +\Delta_0^2}-\xi_\textbf{k}\right)\delta\left(\omega -\sqrt{\xi_\textbf{k}^2 +\Delta_0^2} \right) + \left(\sqrt{\xi_\textbf{k}^2 +\Delta_0^2}+\xi_\textbf{k}\right)\delta\left(\omega +\sqrt{\xi_\textbf{k}^2 +\Delta_0^2} \right)\right] \\
F^{R/A}_0(\textbf{k},\omega)&=\lim_{\eta\rightarrow 0^+}\dfrac{-\Delta_0}{(\omega\pm i\eta)^2-\xi_\textbf{k}^2 -\Delta_0^2} \\
G^{R/A}_0(\textbf{k},\omega)&=\lim_{\eta\rightarrow 0^+}\dfrac{\omega\pm i\eta + \xi_\textbf{k}}{(\omega\pm i\eta)^2-\xi_\textbf{k}^2 -\Delta_0^2} \\
\overline{G}^{R/A}_0(\textbf{k},\omega)&=\lim_{\eta\rightarrow 0^+}\dfrac{\omega\pm i\eta - \xi_\textbf{k}}{(\omega\pm i\eta)^2-\xi_\textbf{k}^2 -\Delta_0^2} \\
\end{aligned}
\label{eq:gfK}
\end{equation}
\end{widetext}
where $\xi_\textbf{k}=\frac{\hbar^2}{2m}k^2 -\mu$ is the single particle dispersion and $\beta$ is the inverse temperature of the system before the drive.
 
Using the expressions in Eqs (\ref{eq:deltaF}) and (\ref{eq:gfK}) one can show that Eq (\ref{eq:gap_linear}) becomes:
\begin{widetext}
\begin{equation}
\begin{aligned}
\delta\Delta(\textbf{r},t)&=\frac{\lambda}{4}\Delta_0\int\frac{d\textbf{q}}{(2\pi)^2}\frac{d\textbf{Q}}{(2\pi)^2}\frac{d\Omega}{2\pi}e^{-i\textbf{Q}\cdot\textbf{r}-i\Omega t}U_{\textbf{Q},\Omega}\left(\Omega + \xi_{\textbf{q}+\tfrac{\textbf{Q}}{2}}+\xi_{\textbf{q}-\tfrac{\textbf{Q}}{2}} \right) \\
&\times \left[  g_{+}(\textbf{Q},\Omega) + g_{-}(\textbf{Q},-\Omega) + g_{-}(-\textbf{Q},\Omega) + g_{+}(-\textbf{Q},-\Omega)  \right]
\end{aligned}
\label{eq:dd}
\end{equation}
\end{widetext}
where we have defined
\begin{equation}
g_{\pm}(\textbf{Q},\Omega)=\frac{1}{E_{\textbf{q}+\tfrac{\textbf{Q}}{2}}}\frac{\tanh\left(\frac{\beta}{2}E_{\textbf{q}+\tfrac{\textbf{Q}}{2}}\right)\pm2}{\Omega^2-2\Omega E_{\textbf{q}+\tfrac{\textbf{Q}}{2}} +\xi_{\textbf{q}+\tfrac{\textbf{Q}}{2}}^2-\xi_{\textbf{q}-\tfrac{\textbf{Q}}{2}}^2 }
\end{equation}
where $E_{\textbf{k}}=\sqrt{\xi_\textbf{k}^2+\Delta_0^2}$. 

To simplify the analysis we do the following: (i) we focus on just one harmonic of the drive potential and assume for the remainder of this section that $U_{\textbf{Q},\Omega}=U_0(2\pi)^3\delta(\textbf{Q}-\textbf{Q}_0)\delta(\Omega-\Omega_0)$; (ii) we assume that the temperature before turning on the drive is much less than the gap allowing us to take $\tanh\left( \frac{\beta}{2}E_{\textbf{q}+\tfrac{\textbf{Q}}{2}}\right)\approx 1$; (iii) we expand Eq (\ref{eq:dd}) in a Taylor series in both $Q_0$ and $\Omega_0$ keeping terms linear in $\Omega_0$ and quadratic in $Q_0$. After these steps Eq (\ref{eq:dd}) becomes:
\begin{widetext}
\begin{equation}
\delta\Delta(\textbf{r},t)=\frac{-\lambda}{4}\Delta_0 U_0 e^{-i\textbf{Q}_0\cdot\textbf{r}-i\Omega_0 t} \int\frac{d\textbf{q}}{(2\pi)^2}\frac{1}{E_{\textbf{q}}^3}\left[\Omega_0 + 2\xi_{\textbf{q}} + \frac{\hbar^2}{2m}Q_0^2\left(\frac{1}{2} -\left[ \frac{3\xi_\textbf{q}^2}{2E_{\textbf{q}}^2} +\xi_\textbf{q}\frac{\hbar^2}{2m}q^2\cos^2{\theta}\left(\frac{5\Delta_0^2}{E_{\textbf{q}}^4}-\frac{2}{E_{\textbf{q}}^2} \right) \right] \right) \right]
\label{eq:ddintegrals}
\end{equation}   
\end{widetext}
where $\theta$ is the angle between $\textbf{Q}_0$ and $\textbf{q}$. 

To evaluate the integrals in Eq (\ref{eq:ddintegrals}) we make use of the assumption that $\Delta_0<<\mu$, which is generally true, and we arrive at the expression:
\begin{equation}
\frac{\delta\Delta(\textbf{r},t)}{\Delta_0}=\frac{-\lambda N_0}{4}\frac{U_0}{\mu} e^{-i\textbf{Q}_0\cdot\textbf{r}-i\Omega_0 t} \left[2+\frac{\Omega_0}{2\mu}+\frac{\pi^2}{12}\overline{Q}_0^2\frac{\Delta_0^2}{\mu^2} \right] 
\label{eq:ddfinal}
\end{equation}
where $\overline{Q}_0$ is the driving wavevector times the coherence length, $\xi=\hbar v_F/\pi\Delta_0$, and $N_0=m/2\pi\hbar^2$ is the density of states. In the text we argue, using parameters for thin films of Pb, that under experimentally realizable conditions $\delta\Delta(\textbf{r},t)$ is negligible.

\bibliographystyle{apsrev}
\bibliography{Odd_Frequency}

\begin{thebibliography}{38}
\expandafter\ifx\csname natexlab\endcsname\relax\def\natexlab#1{#1}\fi
\expandafter\ifx\csname bibnamefont\endcsname\relax
  \def\bibnamefont#1{#1}\fi
\expandafter\ifx\csname bibfnamefont\endcsname\relax
  \def\bibfnamefont#1{#1}\fi
\expandafter\ifx\csname citenamefont\endcsname\relax
  \def\citenamefont#1{#1}\fi
\expandafter\ifx\csname url\endcsname\relax
  \def\url#1{\texttt{#1}}\fi
\expandafter\ifx\csname urlprefix\endcsname\relax\def\urlprefix{URL }\fi
\providecommand{\bibinfo}[2]{#2}
\providecommand{\eprint}[2][]{\url{#2}}

\bibitem[{\citenamefont{Berezinskii}(1974)}]{Berezinskii1974}
\bibinfo{author}{\bibfnamefont{V.~L.} \bibnamefont{Berezinskii}},
  \bibinfo{journal}{Pis' ma Zh. Eksp. Teor. Fiz.}
  \textbf{\bibinfo{volume}{20}}, \bibinfo{pages}{628} (\bibinfo{year}{1974}).

\bibitem[{\citenamefont{Balatsky and Abrahams}(1992)}]{BalatskyPRB1992}
\bibinfo{author}{\bibfnamefont{A.}~\bibnamefont{Balatsky}} \bibnamefont{and}
  \bibinfo{author}{\bibfnamefont{E.}~\bibnamefont{Abrahams}},
  \bibinfo{journal}{Phys. Rev. B} \textbf{\bibinfo{volume}{45}},
  \bibinfo{pages}{13125} (\bibinfo{year}{1992}).

\bibitem[{\citenamefont{Heid}(1995)}]{heid1995thermodynamic}
\bibinfo{author}{\bibfnamefont{R.}~\bibnamefont{Heid}},
  \bibinfo{journal}{Zeitschrift f{\"u}r Physik B Condensed Matter}
  \textbf{\bibinfo{volume}{99}}, \bibinfo{pages}{15} (\bibinfo{year}{1995}).

\bibitem[{\citenamefont{Solenov et~al.}(2009)\citenamefont{Solenov, Martin, and
  Mozyrsky}}]{solenov2009thermodynamical}
\bibinfo{author}{\bibfnamefont{D.}~\bibnamefont{Solenov}},
  \bibinfo{author}{\bibfnamefont{I.}~\bibnamefont{Martin}}, \bibnamefont{and}
  \bibinfo{author}{\bibfnamefont{D.}~\bibnamefont{Mozyrsky}},
  \bibinfo{journal}{Physical Review B} \textbf{\bibinfo{volume}{79}},
  \bibinfo{pages}{132502} (\bibinfo{year}{2009}).

\bibitem[{\citenamefont{Kusunose et~al.}(2011)\citenamefont{Kusunose, Fuseya,
  and Miyake}}]{kusunose2011puzzle}
\bibinfo{author}{\bibfnamefont{H.}~\bibnamefont{Kusunose}},
  \bibinfo{author}{\bibfnamefont{Y.}~\bibnamefont{Fuseya}}, \bibnamefont{and}
  \bibinfo{author}{\bibfnamefont{K.}~\bibnamefont{Miyake}},
  \bibinfo{journal}{Journal of the Physical Society of Japan}
  \textbf{\bibinfo{volume}{80}}, \bibinfo{pages}{054702}
  (\bibinfo{year}{2011}).

\bibitem[{\citenamefont{Bergeret et~al.}(2001)\citenamefont{Bergeret, Volkov,
  and Efetov}}]{BergeretPRL2001}
\bibinfo{author}{\bibfnamefont{F.}~\bibnamefont{Bergeret}},
  \bibinfo{author}{\bibfnamefont{A.}~\bibnamefont{Volkov}}, \bibnamefont{and}
  \bibinfo{author}{\bibfnamefont{K.}~\bibnamefont{Efetov}},
  \bibinfo{journal}{Phys. Rev. Lett.} \textbf{\bibinfo{volume}{86}},
  \bibinfo{pages}{4096} (\bibinfo{year}{2001}).

\bibitem[{\citenamefont{Bergeret et~al.}(2005)\citenamefont{Bergeret, Volkov,
  and Efetov}}]{bergeret2005odd}
\bibinfo{author}{\bibfnamefont{F.}~\bibnamefont{Bergeret}},
  \bibinfo{author}{\bibfnamefont{A.}~\bibnamefont{Volkov}}, \bibnamefont{and}
  \bibinfo{author}{\bibfnamefont{K.}~\bibnamefont{Efetov}},
  \bibinfo{journal}{Reviews of modern physics} \textbf{\bibinfo{volume}{77}},
  \bibinfo{pages}{1321} (\bibinfo{year}{2005}).

\bibitem[{\citenamefont{Yokoyama et~al.}(2007)\citenamefont{Yokoyama, Tanaka,
  and Golubov}}]{yokoyama2007manifestation}
\bibinfo{author}{\bibfnamefont{T.}~\bibnamefont{Yokoyama}},
  \bibinfo{author}{\bibfnamefont{Y.}~\bibnamefont{Tanaka}}, \bibnamefont{and}
  \bibinfo{author}{\bibfnamefont{A.}~\bibnamefont{Golubov}},
  \bibinfo{journal}{Physical Review B} \textbf{\bibinfo{volume}{75}},
  \bibinfo{pages}{134510} (\bibinfo{year}{2007}).

\bibitem[{\citenamefont{Houzet}(2008)}]{houzet2008ferromagnetic}
\bibinfo{author}{\bibfnamefont{M.}~\bibnamefont{Houzet}},
  \bibinfo{journal}{Physical review letters} \textbf{\bibinfo{volume}{101}},
  \bibinfo{pages}{057009} (\bibinfo{year}{2008}).

\bibitem[{\citenamefont{Eschrig and L{\"o}fwander}(2008)}]{EschrigNat2008}
\bibinfo{author}{\bibfnamefont{M.}~\bibnamefont{Eschrig}} \bibnamefont{and}
  \bibinfo{author}{\bibfnamefont{T.}~\bibnamefont{L{\"o}fwander}},
  \bibinfo{journal}{Nature Physics} \textbf{\bibinfo{volume}{4}},
  \bibinfo{pages}{138} (\bibinfo{year}{2008}).

\bibitem[{\citenamefont{Linder et~al.}(2008)\citenamefont{Linder, Yokoyama, and
  Sudb{\o}}}]{LinderPRB2008}
\bibinfo{author}{\bibfnamefont{J.}~\bibnamefont{Linder}},
  \bibinfo{author}{\bibfnamefont{T.}~\bibnamefont{Yokoyama}}, \bibnamefont{and}
  \bibinfo{author}{\bibfnamefont{A.}~\bibnamefont{Sudb{\o}}},
  \bibinfo{journal}{Phys. Rev. B} \textbf{\bibinfo{volume}{77}},
  \bibinfo{pages}{174514} (\bibinfo{year}{2008}).

\bibitem[{\citenamefont{Cr{\'e}pin et~al.}(2015)\citenamefont{Cr{\'e}pin,
  Burset, and Trauzettel}}]{crepin2015odd}
\bibinfo{author}{\bibfnamefont{F.}~\bibnamefont{Cr{\'e}pin}},
  \bibinfo{author}{\bibfnamefont{P.}~\bibnamefont{Burset}}, \bibnamefont{and}
  \bibinfo{author}{\bibfnamefont{B.}~\bibnamefont{Trauzettel}},
  \bibinfo{journal}{Physical Review B} \textbf{\bibinfo{volume}{92}},
  \bibinfo{pages}{100507} (\bibinfo{year}{2015}).

\bibitem[{\citenamefont{Di~Bernardo et~al.}(2015)\citenamefont{Di~Bernardo,
  Diesch, Gu, Linder, Divitini, Ducati, Scheer, Blamire, and
  Robinson}}]{di2015signature}
\bibinfo{author}{\bibfnamefont{A.}~\bibnamefont{Di~Bernardo}},
  \bibinfo{author}{\bibfnamefont{S.}~\bibnamefont{Diesch}},
  \bibinfo{author}{\bibfnamefont{Y.}~\bibnamefont{Gu}},
  \bibinfo{author}{\bibfnamefont{J.}~\bibnamefont{Linder}},
  \bibinfo{author}{\bibfnamefont{G.}~\bibnamefont{Divitini}},
  \bibinfo{author}{\bibfnamefont{C.}~\bibnamefont{Ducati}},
  \bibinfo{author}{\bibfnamefont{E.}~\bibnamefont{Scheer}},
  \bibinfo{author}{\bibfnamefont{M.~G.} \bibnamefont{Blamire}},
  \bibnamefont{and} \bibinfo{author}{\bibfnamefont{J.~W.}
  \bibnamefont{Robinson}}, \bibinfo{journal}{Nature communications}
  \textbf{\bibinfo{volume}{6}} (\bibinfo{year}{2015}).

\bibitem[{\citenamefont{Yokoyama}(2012)}]{YokoyamaPRB2012}
\bibinfo{author}{\bibfnamefont{T.}~\bibnamefont{Yokoyama}},
  \bibinfo{journal}{Phys. Rev. B} \textbf{\bibinfo{volume}{86}},
  \bibinfo{pages}{075410} (\bibinfo{year}{2012}).

\bibitem[{\citenamefont{Black-Schaffer and
  Balatsky}(2012)}]{Black-SchafferPRB2012}
\bibinfo{author}{\bibfnamefont{A.}~\bibnamefont{Black-Schaffer}}
  \bibnamefont{and} \bibinfo{author}{\bibfnamefont{A.}~\bibnamefont{Balatsky}},
  \bibinfo{journal}{Phys. Rev. B} \textbf{\bibinfo{volume}{86}},
  \bibinfo{pages}{144506} (\bibinfo{year}{2012}).

\bibitem[{\citenamefont{Black-Schaffer and
  Balatsky}(2013{\natexlab{a}})}]{Black-SchafferPRB2013}
\bibinfo{author}{\bibfnamefont{A.}~\bibnamefont{Black-Schaffer}}
  \bibnamefont{and} \bibinfo{author}{\bibfnamefont{A.}~\bibnamefont{Balatsky}},
  \bibinfo{journal}{Phys. Rev. B} \textbf{\bibinfo{volume}{87}},
  \bibinfo{pages}{220506(R)} (\bibinfo{year}{2013}{\natexlab{a}}).

\bibitem[{\citenamefont{Triola et~al.}(2014)\citenamefont{Triola, Rossi, and
  Balatsky}}]{TriolaPRB2014}
\bibinfo{author}{\bibfnamefont{C.}~\bibnamefont{Triola}},
  \bibinfo{author}{\bibfnamefont{E.}~\bibnamefont{Rossi}}, \bibnamefont{and}
  \bibinfo{author}{\bibfnamefont{A.~V.} \bibnamefont{Balatsky}},
  \bibinfo{journal}{Phys. Rev. B} \textbf{\bibinfo{volume}{89}},
  \bibinfo{pages}{165309} (\bibinfo{year}{2014}).

\bibitem[{\citenamefont{Tanaka and Golubov}(2007)}]{tanaka2007theory}
\bibinfo{author}{\bibfnamefont{Y.}~\bibnamefont{Tanaka}} \bibnamefont{and}
  \bibinfo{author}{\bibfnamefont{A.}~\bibnamefont{Golubov}},
  \bibinfo{journal}{Physical review letters} \textbf{\bibinfo{volume}{98}},
  \bibinfo{pages}{037003} (\bibinfo{year}{2007}).

\bibitem[{\citenamefont{Tanaka et~al.}(2007)\citenamefont{Tanaka, Tanuma, and
  Golubov}}]{TanakaPRB2007}
\bibinfo{author}{\bibfnamefont{Y.}~\bibnamefont{Tanaka}},
  \bibinfo{author}{\bibfnamefont{Y.}~\bibnamefont{Tanuma}}, \bibnamefont{and}
  \bibinfo{author}{\bibfnamefont{A.}~\bibnamefont{Golubov}},
  \bibinfo{journal}{Phys. Rev. B} \textbf{\bibinfo{volume}{76}},
  \bibinfo{pages}{054522} (\bibinfo{year}{2007}).

\bibitem[{\citenamefont{Linder et~al.}(2009)\citenamefont{Linder, Yokoyama,
  Sudb{\o}, and Eschrig}}]{LinderPRL2009}
\bibinfo{author}{\bibfnamefont{J.}~\bibnamefont{Linder}},
  \bibinfo{author}{\bibfnamefont{T.}~\bibnamefont{Yokoyama}},
  \bibinfo{author}{\bibfnamefont{A.}~\bibnamefont{Sudb{\o}}}, \bibnamefont{and}
  \bibinfo{author}{\bibfnamefont{M.}~\bibnamefont{Eschrig}},
  \bibinfo{journal}{Phys. Rev. Lett.} \textbf{\bibinfo{volume}{102}},
  \bibinfo{pages}{107008} (\bibinfo{year}{2009}).

\bibitem[{\citenamefont{Linder et~al.}(2010)\citenamefont{Linder, Sudb{\o},
  Yokoyama, Grein, and Eschrig}}]{LinderPRB2010_2}
\bibinfo{author}{\bibfnamefont{J.}~\bibnamefont{Linder}},
  \bibinfo{author}{\bibfnamefont{A.}~\bibnamefont{Sudb{\o}}},
  \bibinfo{author}{\bibfnamefont{T.}~\bibnamefont{Yokoyama}},
  \bibinfo{author}{\bibfnamefont{R.}~\bibnamefont{Grein}}, \bibnamefont{and}
  \bibinfo{author}{\bibfnamefont{M.}~\bibnamefont{Eschrig}},
  \bibinfo{journal}{Phys. Rev. B} \textbf{\bibinfo{volume}{81}},
  \bibinfo{pages}{214504} (\bibinfo{year}{2010}).

\bibitem[{\citenamefont{Tanaka et~al.}(2012)\citenamefont{Tanaka, Sato, and
  Nagaosa}}]{TanakaJPSJ2012}
\bibinfo{author}{\bibfnamefont{Y.}~\bibnamefont{Tanaka}},
  \bibinfo{author}{\bibfnamefont{M.}~\bibnamefont{Sato}}, \bibnamefont{and}
  \bibinfo{author}{\bibfnamefont{N.}~\bibnamefont{Nagaosa}},
  \bibinfo{journal}{J. Phys. Soc. Jpn.} \textbf{\bibinfo{volume}{81}},
  \bibinfo{pages}{011013} (\bibinfo{year}{2012}).

\bibitem[{\citenamefont{Black-Schaffer and
  Balatsky}(2013{\natexlab{b}})}]{black2013odd}
\bibinfo{author}{\bibfnamefont{A.~M.} \bibnamefont{Black-Schaffer}}
  \bibnamefont{and} \bibinfo{author}{\bibfnamefont{A.~V.}
  \bibnamefont{Balatsky}}, \bibinfo{journal}{Physical Review B}
  \textbf{\bibinfo{volume}{88}}, \bibinfo{pages}{104514}
  (\bibinfo{year}{2013}{\natexlab{b}}).

\bibitem[{\citenamefont{Komendova et~al.}(2015)\citenamefont{Komendova,
  Balatsky, and Black-Schaffer}}]{komendova2015experimentally}
\bibinfo{author}{\bibfnamefont{L.}~\bibnamefont{Komendova}},
  \bibinfo{author}{\bibfnamefont{A.~V.} \bibnamefont{Balatsky}},
  \bibnamefont{and} \bibinfo{author}{\bibfnamefont{A.~M.}
  \bibnamefont{Black-Schaffer}}, \bibinfo{journal}{Physical Review B}
  \textbf{\bibinfo{volume}{92}}, \bibinfo{pages}{094517}
  (\bibinfo{year}{2015}).

\bibitem[{\citenamefont{Parhizgar and
  Black-Schaffer}(2014)}]{parhizgar_2014_prb}
\bibinfo{author}{\bibfnamefont{F.}~\bibnamefont{Parhizgar}} \bibnamefont{and}
  \bibinfo{author}{\bibfnamefont{A.~M.} \bibnamefont{Black-Schaffer}},
  \bibinfo{journal}{Physical Review B} \textbf{\bibinfo{volume}{90}},
  \bibinfo{pages}{184517} (\bibinfo{year}{2014}).

\bibitem[{\citenamefont{Triola et~al.}(2016)\citenamefont{Triola, Badiane,
  Balatsky, and Rossi}}]{triola2016}
\bibinfo{author}{\bibfnamefont{C.}~\bibnamefont{Triola}},
  \bibinfo{author}{\bibfnamefont{D.~M.} \bibnamefont{Badiane}},
  \bibinfo{author}{\bibfnamefont{A.~V.} \bibnamefont{Balatsky}},
  \bibnamefont{and} \bibinfo{author}{\bibfnamefont{E.}~\bibnamefont{Rossi}},
  \bibinfo{journal}{Physical Review Letters} \textbf{\bibinfo{volume}{116}},
  \bibinfo{pages}{257001} (\bibinfo{year}{2016}).

\bibitem[{\citenamefont{Pivovarov and Nayak}(2001)}]{pivovarov2001odd}
\bibinfo{author}{\bibfnamefont{E.}~\bibnamefont{Pivovarov}} \bibnamefont{and}
  \bibinfo{author}{\bibfnamefont{C.}~\bibnamefont{Nayak}},
  \bibinfo{journal}{Physical Review B} \textbf{\bibinfo{volume}{64}},
  \bibinfo{pages}{035107} (\bibinfo{year}{2001}).

\bibitem[{\citenamefont{Kedem and Balatsky}(2015)}]{kedem2015odd}
\bibinfo{author}{\bibfnamefont{Y.}~\bibnamefont{Kedem}} \bibnamefont{and}
  \bibinfo{author}{\bibfnamefont{A.~V.} \bibnamefont{Balatsky}},
  \bibinfo{journal}{arXiv preprint arXiv:1501.07049}  (\bibinfo{year}{2015}).

\bibitem[{\citenamefont{Huang et~al.}(2015)\citenamefont{Huang, W{\"o}lfle, and
  Balatsky}}]{huang2015odd}
\bibinfo{author}{\bibfnamefont{Z.}~\bibnamefont{Huang}},
  \bibinfo{author}{\bibfnamefont{P.}~\bibnamefont{W{\"o}lfle}},
  \bibnamefont{and} \bibinfo{author}{\bibfnamefont{A.}~\bibnamefont{Balatsky}},
  \bibinfo{journal}{Physical Review B} \textbf{\bibinfo{volume}{92}},
  \bibinfo{pages}{121404} (\bibinfo{year}{2015}).

\bibitem[{\citenamefont{Zhang et~al.}(2005)\citenamefont{Zhang, Jia, Han, Tang,
  Shen, Guo, Qiu, and Xue}}]{zhang2005band}
\bibinfo{author}{\bibfnamefont{Y.-F.} \bibnamefont{Zhang}},
  \bibinfo{author}{\bibfnamefont{J.-F.} \bibnamefont{Jia}},
  \bibinfo{author}{\bibfnamefont{T.-Z.} \bibnamefont{Han}},
  \bibinfo{author}{\bibfnamefont{Z.}~\bibnamefont{Tang}},
  \bibinfo{author}{\bibfnamefont{Q.-T.} \bibnamefont{Shen}},
  \bibinfo{author}{\bibfnamefont{Y.}~\bibnamefont{Guo}},
  \bibinfo{author}{\bibfnamefont{Z.}~\bibnamefont{Qiu}}, \bibnamefont{and}
  \bibinfo{author}{\bibfnamefont{Q.-K.} \bibnamefont{Xue}},
  \bibinfo{journal}{Physical review letters} \textbf{\bibinfo{volume}{95}},
  \bibinfo{pages}{096802} (\bibinfo{year}{2005}).

\bibitem[{\citenamefont{Eom et~al.}(2006)\citenamefont{Eom, Qin, Chou, and
  Shih}}]{eom2006persistent}
\bibinfo{author}{\bibfnamefont{D.}~\bibnamefont{Eom}},
  \bibinfo{author}{\bibfnamefont{S.}~\bibnamefont{Qin}},
  \bibinfo{author}{\bibfnamefont{M.-Y.} \bibnamefont{Chou}}, \bibnamefont{and}
  \bibinfo{author}{\bibfnamefont{C.}~\bibnamefont{Shih}},
  \bibinfo{journal}{Physical review letters} \textbf{\bibinfo{volume}{96}},
  \bibinfo{pages}{027005} (\bibinfo{year}{2006}).

\bibitem[{\citenamefont{Brun et~al.}(2009)\citenamefont{Brun, Hong, Patthey,
  Sklyadneva, Heid, Echenique, Bohnen, Chulkov, and
  Schneider}}]{brun2009reduction}
\bibinfo{author}{\bibfnamefont{C.}~\bibnamefont{Brun}},
  \bibinfo{author}{\bibfnamefont{I.-P.} \bibnamefont{Hong}},
  \bibinfo{author}{\bibfnamefont{F.}~\bibnamefont{Patthey}},
  \bibinfo{author}{\bibfnamefont{I.~Y.} \bibnamefont{Sklyadneva}},
  \bibinfo{author}{\bibfnamefont{R.}~\bibnamefont{Heid}},
  \bibinfo{author}{\bibfnamefont{P.}~\bibnamefont{Echenique}},
  \bibinfo{author}{\bibfnamefont{K.}~\bibnamefont{Bohnen}},
  \bibinfo{author}{\bibfnamefont{E.}~\bibnamefont{Chulkov}}, \bibnamefont{and}
  \bibinfo{author}{\bibfnamefont{W.-D.} \bibnamefont{Schneider}},
  \bibinfo{journal}{Physical review letters} \textbf{\bibinfo{volume}{102}},
  \bibinfo{pages}{207002} (\bibinfo{year}{2009}).

\bibitem[{\citenamefont{Zhang et~al.}(2010)\citenamefont{Zhang, Cheng, Li, Sun,
  Wang, Zhu, He, Wang, Ma, Chen et~al.}}]{zhang2010superconductivity}
\bibinfo{author}{\bibfnamefont{T.}~\bibnamefont{Zhang}},
  \bibinfo{author}{\bibfnamefont{P.}~\bibnamefont{Cheng}},
  \bibinfo{author}{\bibfnamefont{W.-J.} \bibnamefont{Li}},
  \bibinfo{author}{\bibfnamefont{Y.-J.} \bibnamefont{Sun}},
  \bibinfo{author}{\bibfnamefont{G.}~\bibnamefont{Wang}},
  \bibinfo{author}{\bibfnamefont{X.-G.} \bibnamefont{Zhu}},
  \bibinfo{author}{\bibfnamefont{K.}~\bibnamefont{He}},
  \bibinfo{author}{\bibfnamefont{L.}~\bibnamefont{Wang}},
  \bibinfo{author}{\bibfnamefont{X.}~\bibnamefont{Ma}},
  \bibinfo{author}{\bibfnamefont{X.}~\bibnamefont{Chen}}, \bibnamefont{et~al.},
  \bibinfo{journal}{Nature Physics} \textbf{\bibinfo{volume}{6}},
  \bibinfo{pages}{104} (\bibinfo{year}{2010}).

\bibitem[{\citenamefont{Rammer}(2007)}]{rammer2007quantum}
\bibinfo{author}{\bibfnamefont{J.}~\bibnamefont{Rammer}},
  \emph{\bibinfo{title}{Quantum field theory of non-equilibrium states}}
  (\bibinfo{publisher}{Cambridge University Press}, \bibinfo{year}{2007}).

\bibitem[{\citenamefont{Stefanucci and van
  Leeuwen}(2013)}]{stefanucci2013nonequilibrium}
\bibinfo{author}{\bibfnamefont{G.}~\bibnamefont{Stefanucci}} \bibnamefont{and}
  \bibinfo{author}{\bibfnamefont{R.}~\bibnamefont{van Leeuwen}},
  \emph{\bibinfo{title}{Nonequilibrium Many-Body Theory of Quantum Systems: A
  Modern Introduction}} (\bibinfo{publisher}{Cambridge University Press},
  \bibinfo{year}{2013}).

\bibitem[{\citenamefont{Damascelli et~al.}(2003)\citenamefont{Damascelli,
  Hussain, and Shen}}]{damascelli2003angle}
\bibinfo{author}{\bibfnamefont{A.}~\bibnamefont{Damascelli}},
  \bibinfo{author}{\bibfnamefont{Z.}~\bibnamefont{Hussain}}, \bibnamefont{and}
  \bibinfo{author}{\bibfnamefont{Z.-X.} \bibnamefont{Shen}},
  \bibinfo{journal}{Reviews of modern physics} \textbf{\bibinfo{volume}{75}},
  \bibinfo{pages}{473} (\bibinfo{year}{2003}).

\bibitem[{\citenamefont{Hofer et~al.}(2003)\citenamefont{Hofer, Foster, and
  Shluger}}]{hofer2003theories}
\bibinfo{author}{\bibfnamefont{W.~A.} \bibnamefont{Hofer}},
  \bibinfo{author}{\bibfnamefont{A.~S.} \bibnamefont{Foster}},
  \bibnamefont{and} \bibinfo{author}{\bibfnamefont{A.~L.}
  \bibnamefont{Shluger}}, \bibinfo{journal}{Reviews of Modern Physics}
  \textbf{\bibinfo{volume}{75}}, \bibinfo{pages}{1287} (\bibinfo{year}{2003}).

\bibitem[{\citenamefont{Tersoff and Hamann}(1983)}]{tersoff1983theory}
\bibinfo{author}{\bibfnamefont{J.}~\bibnamefont{Tersoff}} \bibnamefont{and}
  \bibinfo{author}{\bibfnamefont{D.}~\bibnamefont{Hamann}},
  \bibinfo{journal}{Physical review letters} \textbf{\bibinfo{volume}{50}},
  \bibinfo{pages}{1998} (\bibinfo{year}{1983}).

\end{thebibliography}

\end{document}